# Single atom extraction by scanning tunneling microscope tip-crash and nanoscale surface engineering

*Saw -Wai Hla[*], Kai-Felix Braun, Violeta Iancu, Aparna Deshpande*

Nanoscale & Quantum Phenomena Institute, Physics & Astronomy Department,

Ohio University, Athens, OH 45701, USA.

[*]Corresponding author. E-mail: hla@ohio.edu, Web: www.phy.ohiou.edu/~hla

We report a novel atom extraction mechanism from the native substrate by means of a scanning tunneling microscope tip-crash on a Ag(111) surface at 5 K. Individual atoms are scattered on the surface when a silver coated tip is dipped into the substrate at low tunneling biases. Quantitative analyses reveal that the mechanical energy supplied by the tip-crash dominates the atom extraction process. Application of this procedure is demonstrated by constructing quantum structures using the extracted atoms on an atom-by-atom basis.

When a scanning tunneling microscope (STM) tip plunges into a material surface, which is often known as 'tip-crash', it causes a crater (indentation) on the surface. A 'Tip-crash' is an undesired event during STM imaging because it can destroy the surface area under investigation. It can also alter the shape of the tip-apex and thus, many experimentalists prefer to avoid an accidental tip-crash. But the tip-crash can be useful in some cases, e.g. to test the hardness of materials on surfaces[1-5] or to form a nanowire between the tip and sample[6,7]. The progress of the bottom-up approach in nanoscience is further accelerated by the recent invention of the multiple atomic force microscope (AFM) tip-arrays, known as 'millipede'[8], which has over a thousand tips assembled inside a microchip. The application of millipede is based on a similar procedure as the STM tip-crash. The millipede can be used to form multiple indentations on surfaces, which opens up the possibility of using the tip-crash procedure in industrial scale nanofabrications. Further investigation of the tip-crash mechanism is desirable and new finding in this area may have a great impact on the progress of nanoscience. In this letter, we describe a novel application of the tip-crash procedure: Extraction of individual atoms from a native silver substrate.

The experiments were performed by using a homebuilt Besocke-Beetle type ultra-high-vacuum low-temperature scanning tunneling microscope operated at pressures below $4 \times 10^{-11}$ Torr. An electrochemically etched polycrystalline tungsten wire was used for the STM tip. A single crystal Ag(111) sample was cleaned by repeated cycles of neon ion sputtering and annealing to 800 K. The sample was then cooled down to 5 K inside the STM system using a liquid helium bath cryostat. Cleanliness of the Ag(111) surface was checked by STM imaging prior to the tip-crash experiments.

The first experimental step is to coat the tungsten tip-apex with silver, which is realized by a controlled tip-crash into the Ag(111) surface under a high tunneling bias condition. In this procedure, the tip is initially positioned above an atomically flat surface area, and then is dipped into the substrate for a few nm. The coating process has been successfully performed with the tip dipping speeds between 0.2 and 2.5 nm s$^{-1}$ and the applied tunneling biases between +3 and +8 V. The tip penetration depth into the substrate is precisely controlled by selecting a suitable 'z' piezo voltage.

The STM image of Fig. 1A shows the result of an initial tip-crash on the Ag(111) surface where a 3.5 nm deep indentation with a large entrance area (~ 15 nm diameter) has been produced. After the tip-crash, some of the substrate material is piled up as monolayer-high silver islands on the surface next to the indented area. It is known that a mass transport of silver occurs above room temperature resulting in formation of silver islands with well-defined shapes[9]. In our experiment, the piled up materials form islands with straight step-edges indicating (Fig. 1A) that the amount of heat locally produced in the tip-crash is sufficient for the diffusion of silver on the surface. The volume of the indentation is determined from the STM image (Fig. 1A), and it is compared with the volume of the piled up silver islands. The measured volume of the indentation, 206 nm$^3$, is larger than the volume of the piled-up materials, 163 nm$^3$. Here the missing silver volume is attributed as being picked up by the tip. Therefore, the tip is assumed coated with silver after the tip-crash.

After coating the tip-apex with silver, we perform another tip-crash at a location below the previous crash site (Fig. 1B). The applied tunneling voltage for this second tip-crash is set at 1.5 V and the tip penetration depth is precisely controlled to be 3.5 nm, which is the same depth as before. An interesting result here is that the STM line profile across the new indentation shows a much smaller entrance area (~ 10 nm) than the previous one, and the indented area has a sharp triangular profile (Fig. 1B). A comparison between the indentation line profiles shows that the tip-apex, which is already covered by silver, is also reformed after the previous tip-crash and has now a more symmetric shape. In most cases, the STM image contrast is also enhanced after the tip-crash indicating that the tip becomes sharper. Hence, a sharp tip-apex with a known chemical identity can be formed locally using this procedure.

The scattered atoms observed in the STM image after the tip-crash can be either tungsten atoms or silver atoms. In order to verify the chemical identity of the extracted atoms, STM lateral manipulation (LM) experiments have been performed. A sequence of STM images in Fig. 2 shows a manipulation scheme used here to identify the chemical identity of the atoms. In the first STM image (Fig. 2A), several extracted atoms (white circles) and small holes (dark areas) produced by tip-crashes can be observed. A small hole at the middle of this image measured as 0.8 nm wide and 0.07 nm deep, which is indicated with a dashed circle, is filled with the extracted atoms by manipulating them with the STM tip on an atom-by-atom basis[10,11]. The sequence of STM images (Fig. 2A to 2F) shows a gradual filling of the hole during the process. After the fourth atom filling, the hole completely disappears from the surface. This can be observed in Fig. 2F where the former hole area, which is indicated with a dashed circle, is not distinguishable from the surrounding surface anymore. On metallic surfaces, different types of atoms cause an image contrast due to electronic structural differences between the chemical elements[12]. Since no image contrast is produced at the hole area after filling, we conclude that the extracted atoms are indeed silver atoms. Repeating of this manipulation scheme provides similar results for the atoms extracted with a bias range between +1.5 V and +4 V.

To study the mechanism of atom extraction, the procedure has been repeated by varying the tunneling voltage between +1.5 to +4 V and the tip-penetration depth between 0.2 to 2.5 nm. The tunneling currents at initial tip positions are varied between 0.5 to 2 nA. We find that the displacement of scattered atoms from the tip-crash site increases with increasing tip-sample contact area (Fig. 3A). Here the tip apex is assumed as a cone shape and the contact area is determined by calculating the surface area of a cone; $Area = \pi\left(r\sqrt{r^2 + d^2}\right)$, where '$r$' and '$d$' are the radius and depth of the indentation. Both values are determined from the STM images. Statistical analysis reveals that the initial tunneling current settings do not play a significant role in the scattered atom distance. However, the current assisted detachment of the atoms from the tip or surface cannot be ruled out[13]. There is a linear dependency of the displacement of the atoms away from the tip-crash location upon the applied tip-sample junction biases. Fig. 3B and 3C present the plots of atom displacement as a function of applied bias and as a function of normalized tip-surface contact area by means of bias. There is a weak linear increase in Fig. 3B indicating that the electric field effect slightly contributes to the process. However, the curve in Fig. 3C, which is the tip-sample contact area normalized by means of bias vs displacement, shows a similar distribution profile as in Fig. 3A, which is without normalization with respect to bias. This indicates that the tip-sample mechanical contact is the main driving force in our atom extraction process.

For a silver atom to be located some distance away from a tip-crash site, it needs to be transported from the indented area. There are two possible mechanisms, which shall be discussed here: (1) The atom diffuses across the surface after detachment from the tip-surface junction area. (2) The atom is ejected due to the impact of tip-crash or due to the breaking of a nanowire formed between the tip and sample[6,7], and later re-adsorbs at a surface site. The atom diffusion process might be fueled by current induced excitations[14], electric field effect and/or surface phonons[13]. For both cases, the tip apex needs to be located at the proximity of the atom if low tunneling biases are used[13]. In the STM images acquired after the tip-crash, the scattered atoms can be observed as far as ~180 nm from the crash site. This distance is further away from the tip-apex location and the current or electric field effect induced hopping of the atoms will become negligible. In addition, the length of piled up islands with straight step-edges can be up to 40 nm. Such island formation, which requires a higher voltage range than the atom extraction process, is assumed as caused by thermal diffusion of silver. This length is much shorter than the maximum atom displacement of ~180 nm, which involves a lower voltage range, and hence we assume that the phonon contribution in the silver diffusion process is not significant. Thus the diffusion of silver atoms across the Ag(111) surface may not be the only cause.

In order to further clarify the mechanism of extraction, we have performed tip-crashes at surface step areas. Fig. 3D shows an STM image taken after a tip-crash at a step terrace where the scattered atoms are observed at several higher and lower step terraces away from the tip-crash site. Due to the existence of the Ehrlich-Schwoebel barrier[15] at the step, diffusion of silver atoms across several step edges on the Ag(111) surface at this low temperature is unlikely to occur. This is further supported by the STM tip induced diffusion of single silver atoms across the surface using the STM lateral manipulation procedure[10,11]. We find that individual silver atoms can be moved on the surface anywhere in the flat terrace area however the atoms can not be transported across the Ag(111) step-edge using lateral manipulation. Attempts to move the atom across step edges always result in binding of the atoms at the lower part of the step edge. Thus, the observed scattered atoms at several step terraces higher than the tip-crash site have to be transported by ejecting them from the tip-crash location.

During the construction of quantum corrals and atomic scale structures on surfaces using STM manipulation, the basic building block of construction materials --individual atoms-- are usually deposited on the surface prior to the experiment[16,17]. We now have a way to locally produce individual atoms from the native substrate using the tip-crash procedure. As a demonstration, the quantum structures (Fig. 4) are constructed using these extracted atoms. Fig 4A shows an STM image acquired at a region next to an indented area illustrated in Fig. 1 where the islands formed by the pile-up substrate materials are recognizable. The scattered individual silver atoms observed in the image are produced by the second tip-crash discussed at the tip forming section above. By using the STM lateral manipulation procedure[10,11], the atoms are relocated to form two one-dimensional atomic arrays (Fig. 4B and 4C) with the tip. This 'electron resonator' is designed to have a linear standing wave front for molecular diffusion experiments. Fig. 4D illustrates an example of atomistic art, "atomic smiley" image, constructed with native silver atoms extracted by using the tip-crash procedure. An STM movie showing the smiley construction process is provided in the supplemental materials section.

In conclusion, we have demonstrated a novel atom extraction scheme by means of a controlled STM tip-crash on a Ag(111) surface. The chemical identity verification of the extracted atoms by filling them back into a hole is a new example of STM manipulation. In addition, the in-situ tip forming procedure described here allows fast and easy tip preparation with known chemical constitution for STM imaging and spectroscopy experiments. Our atom extraction scheme allows construction of atomic scale quantum structures and atomistic devices using native substrate materials locally within an area of a few tens of nm$^2$, in accordance with future nanoscience and nanotechnology applications.

**Acknowledgement:** The financial support provided by the United State Department of Energy, Basic Energy Sciences grant no. DE-FG02-02ER46012 is gratefully acknowledged.

**Figure Captions**

**Figure 1.** STM images of tip indentations. (A) An STM image of Ag(111) surface after an initial tip-crash using 3V tip-sample junction bias, in which the piled up materials form mostly mono-layer high silver islands with straight step-edges. The diameter and depth of indentation are determined from the STM line-profile shown below the image. (B) The STM image acquired after the second tip-crash performed by using 1.5 V tip-sample junction bias shows a smaller indentation entrance. The comparison between the two line profiles reveals that the second tip-crash produces an indentation with a sharper triangular shape and a smaller entrance diameter (10 nm). [Imaging parameters: $V_t = 0.29$ V, $I_t = 1.7$ nA, scan area = 120 x 110 $nm^2$].

**Figure 2.** Hole filling. A sequence of STM images demonstrates atom-by-atom filling of a hole on the surface indicated by a dashed circle at the middle of the image in Fig. 2A. This hole is created by a controlled tip-crash. The white circles are individual atoms, which are scattered after a tip-crash. The red arrows indicate the STM manipulation directions of the atoms during the filling process. The hole completely disappears after filling with the fourth atom. The former hole location is indicated by a dashed circle in Fig. 2F, [Image size: 9 x 6 $nm^2$].

**Figure 3.** Atom extraction mechanism. The tip-sample contact area vs. maximum atom displacement from the tip-crash site (A). The tunneling voltage vs. atom displacement plot with a linear line fit (dashed line) (B). The normalized tip-sample contact area (area/V) vs. atom displacement plot (C). An STM image showing scattered atoms at step terraces after a tip-crash (D). [Image size: 70 x 70 nm$^2$]

**Figure 4.** Atom-by-atom construction of quantum structures. An STM image acquired at an area next to the tip forming site illustrated in Fig. 1A shows scattered silver atoms produced by a tip-crash (A). These atoms are laterally manipulated with the STM tip to construct an electron resonator having two atomic arrays (B) [Image size: 63 x 63 nm$^2$]. A three dimensional representation of the constructed structure is shown in (C) [The resonator dimension: length = 22 nm, width = 8 nm]. A smiley image has been constructed by using the extracted atoms by means of a tip-crash (D) [Image diameter: 32 nm]. The STM movie of this smiley construction is provided in supplementary section.

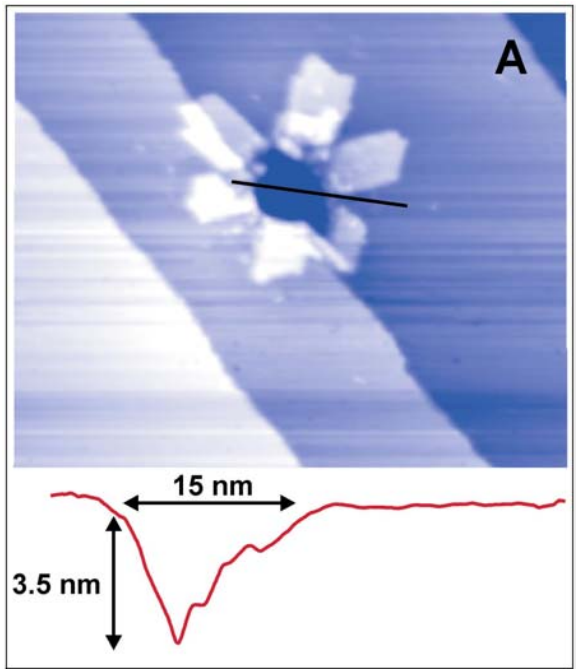

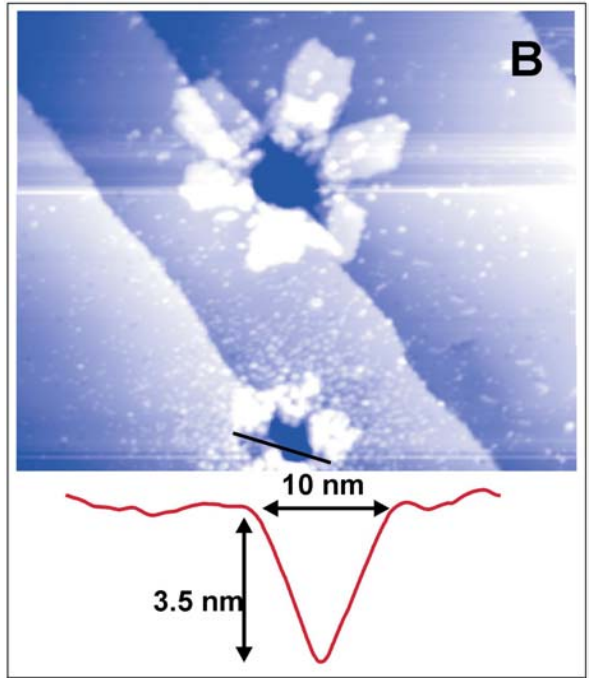

**Figure 1**

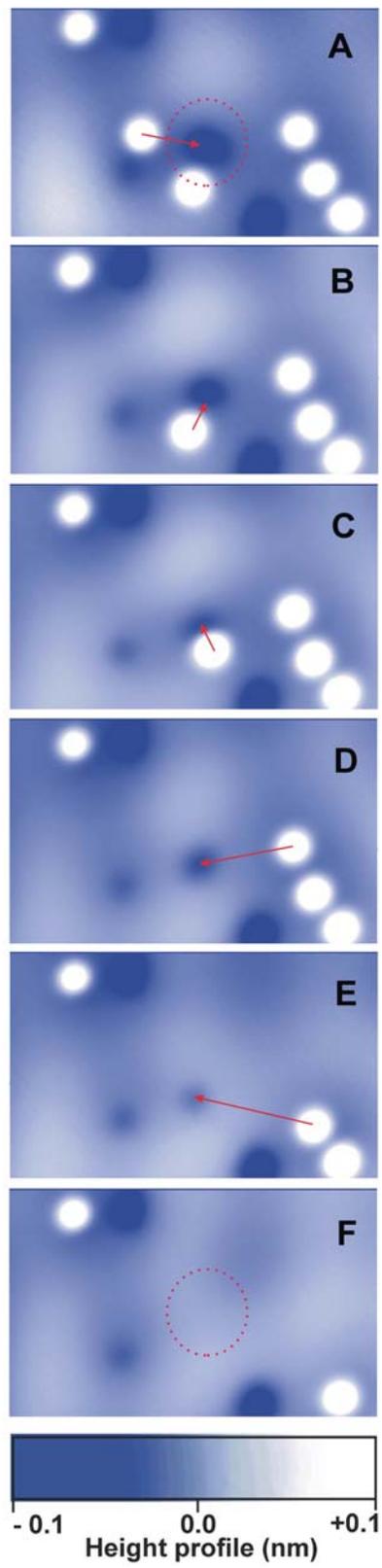

**Figure 2**

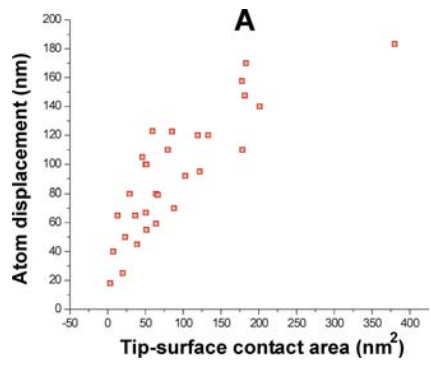
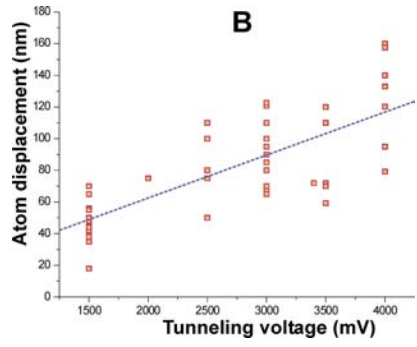
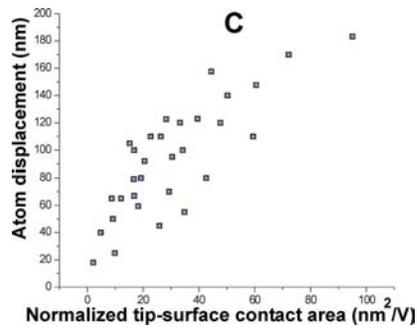
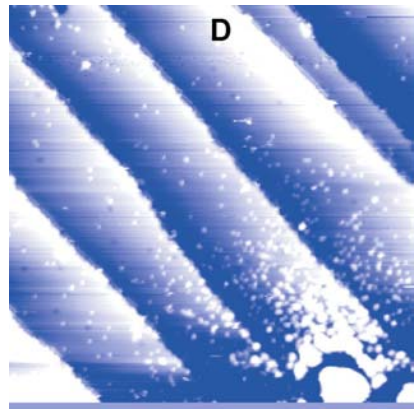

**Figure 3**

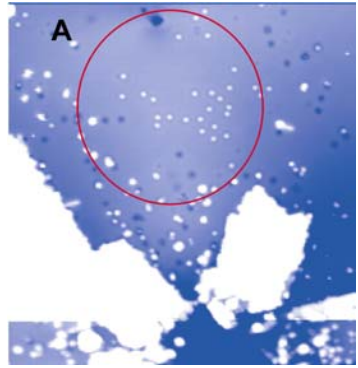

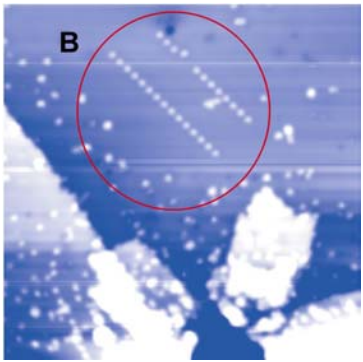

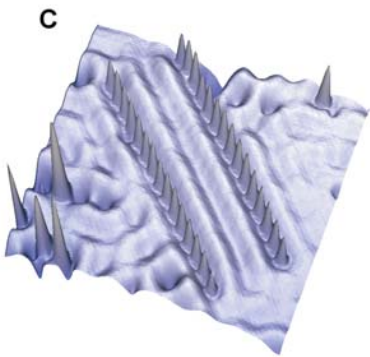

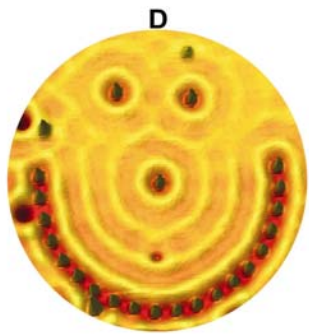

**Figure 4**

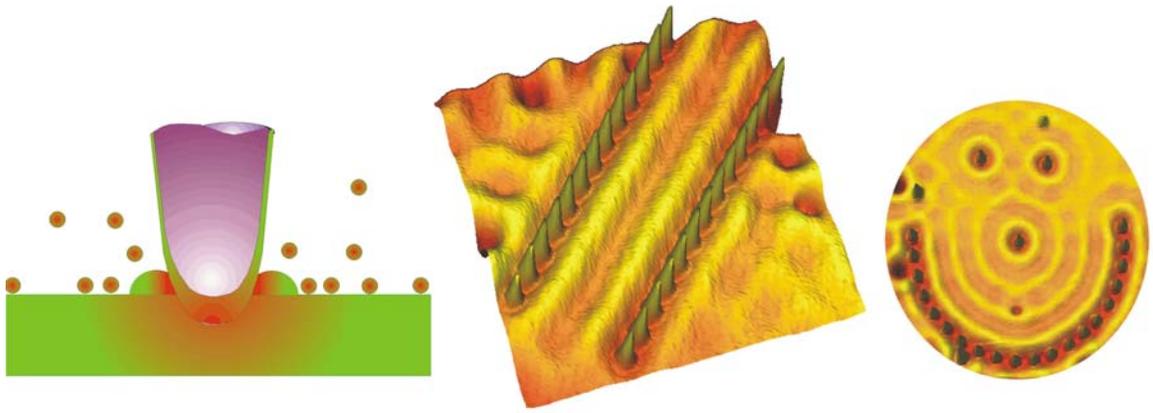

**Title image**